\documentclass[aps,twocolumn,prl,endfloats]{revtex4}
\usepackage{bm}
\usepackage{epsfig}

\begin{document}
\title{Pulsed light beams in vacuum with superluminal and negative group velocities}
\author{Miguel A. Porras}
\email{porras@dfarn.upm.es}
\affiliation{Departamento de F\'\i{}sica Aplicada.
Universidad Polit\'ecnica de Madrid. Rios Rosas 21. E-28003 Madrid. Spain}
\author{Isabel Gonzalo}
\affiliation{Departamento de \'Optica. Universidad Complutense de Madrid. Ciudad Universitaria s/n. E-28040 Madrid. Spain}
\author{Alessia Mondello}
\affiliation{Dipartimento di Fisica. Universit\'a degli Studi Roma Tre. Via della Vasca Navale 84. I-00146 Rome. Italy}

\begin{abstract}
Gouy's phase of transversally limited pulses can create a strong anomalous dispersion in vacuum leading to highly superluminal and negative group velocities. As a consequence,
a focusing pulse can diverge beyond the focus before converging into it. A simple experiment is proposed.
\end{abstract}

\pacs{42.65.Re, 42.65.Tg}

\maketitle

%%%%%%%%%%%%%%%% INTRODUCTION %%%%%%%%%%%%%%%%%%%%%

Propagation of light pulses at superluminal velocities has received a good deal of attention in recent years \cite{CHIAO,DU,WANG}.
Superluminal group velocities in light pulses have been predicted and experimentally demonstrated in evanescent modes of undersized waveguides \cite{CHIAO}, in $X$-Besssel waves \cite{DU}, and in pulses travelling in transparent materials with pronounced enough anomalous dispersion \cite{WANG}, one of the most striking results being a negative superluminal group velocity of an undeformable pulse, which in practice means that the pulse exits the dispersive material before entering \cite{WANG}.

In this letter we show that arbitrarily high and negative group velocities in temporally undeformable pulses are also possible in vacuum as an effect of the phase anomaly \cite{BO}, or Gouy's phase associated to diffraction.

In fact, slight superluminal behavior in collimated \cite{POPRE} and focused \cite{ZBOR} pulsed Gaussian beam has been recently described. Though these results were obtained under the approximate paraxial theory of light beam propagation \cite{SI}, they have been shown to be correct and accurate from the nonparaxial vectorial Kirchhoff-Sommerfeld diffraction formula \cite{POOC1,ZBOR}.

Discrepancy of the group velocity from $c$ in arbitrary pulsed light beams is shown here to be a  consequence of the frequency dependence of Gouy's phase, a dispersion which can be controlled by properly selecting the pulse transversal profile.
We then find it rather simple, by using standard optics such as a lens and a radially graded mirror \cite{BE02,PI96}, to generate a strong enough anomalous dispersion so as to produce arbitrarily high superluminal and negative group velocities in a focusing pulse, whose temporal form is moreover found to remain nearly unchanged.

%%%%%%%%%%%%%%% GENERAL THEORY %%%%%%%%%%%%%%%%%%%%

Let us start by considering the three-dimensional wave packet, or pulsed light beam
\begin{equation}\label{UNO}
E(r,z,t) =\frac{1}{\pi}\int_0^\infty d\omega \, \hat E_\omega(r,z)\exp(-i\omega t) ,
\end{equation}
$r=(x^2+y^2)^{1/2}$, that results from the superposition of monochromatic light beams $\hat E_\omega (r,z)$ of different frequencies $\omega$.

%%%%%%%%%%%%%%%%%  beams %%%%%%%%%%%%%%%%%%%%%

All components $\hat E_\omega (r,z)$ are assumed to be paraxial about the positive $z$-axis,  and further to have, for simplicity, revolution symmetry around this axis.
The slowly varying complex amplitude $\hat \psi_\omega (r,z)$ of the paraxial light beam $\hat E_\omega(r,z)$ [i.e., $\hat E_\omega=\hat\psi_\omega\exp(i\omega z/c)$, with $c$ the vacuum light speed of plane waves]  obeys the paraxial wave equation $\Delta_\perp \hat \psi_\omega+2i(\omega/c)\partial_z \hat \psi_\omega = 0$,
where $\Delta_\perp=\partial_{xx}+ \partial_{yy}=(1/r)\partial_r(r\partial_r)$ is the transversal Laplacian. Writing $\hat \psi_\omega= a_\omega \exp(i\phi_\omega)$, i.e., in terms of  its real amplitude $a_\omega(r,z)=\sqrt{I_\omega(r,z)}>0$
[squared root of intensity $I_\omega(r,z)$] and phase $\phi_\omega(r,z)$, it is a straighforward calculation to obtain the propagation equation for the on-axis ($r=0$) phase $\phi_\omega(z)\equiv\phi_\omega(0,z)$, or Gouy's phase, as $\partial_z\phi_\omega (z)=(c/2\omega)C_\omega (z)$, where $C_\omega(z)\equiv\Delta_\perp a_\omega(r,z)|_{r=0}/a_\omega(0,z)$, and we have assumed that the light beam is smooth enough at on-axis points [$\partial_r\phi(r,z)|_{r=0} =0$].
The axial variation of the Gouy's phase $\partial_z \phi(z)$ of a monochromatic light beam can thus be inferred from the property of the transversal intensity profile $C_\omega(z)$, an easily measurable quantity having an intuitive meaning:
If the normalized intensity profile is expanded in power series about $r=0$, we obtain $I_\omega(r,z)/I_\omega(0,z)=1+ (1/2)C_\omega(z)r^2+\dots$; thus $C_\omega(z)$ gives the concavity of the intensity profile at $r=0$.  For example, $C_\omega(z)=-4/s_\omega^2(z)$ for the Gaussian profile $a_\omega(r,z)=\exp[-r^2/s_\omega^2(z)]$.

%%%%%%%%%%%%%%%% Pulses %%%%%%%%%%%%%%%%%%%%%

If our wavepacket $E(r,z,t)$ has the form of enveloped oscillations of a certain carrier frequency $\omega_0$, then $a_\omega (z)$ is a narrow function of frequency around $\omega_0$  ($\Delta\omega/\omega_0\ll 1$, $\Delta\omega$ being, e.g., the half-width at $1/e$ maximum amplitude). The phase and group velocities at which the carrier oscillations and the envelope propagate in vacuum are given, respectively, by \cite{BO} $v_p=\omega_0/\partial_z[\omega_0 z/c +\phi_{\omega_0}(z)]$ and $v_g=1/\partial_z[\omega z/c+\phi_\omega(z)]'_{\omega_0}$, where the prime sign means differentiation with respect to $\omega$. It is then seen that discrepancies of $v_p$ and $v_g$ from $c$ are due, respectively, to axial and frequency dependence of Gouy's phase, dependences that originate, also respectively, from transversal limitation of the wave (i.e., from diffraction) and the frequency-dependent nature of the diffraction phenomenon.
Negative group velocities (with the physical sense discussed by Wang {\em et al.} \cite{WANG}) are in principle allowed by omitting the absolute value in the rigorous definition of group velocity \cite{BO}.
Since $\partial_z\phi(z)=(c/2\omega)C_\omega(z)$, the alternative expressions
$v_p/c=\left[1+(c^2/2\omega_0^2)C_{\omega_0}(z)\right]^{-1}$, and
\begin{equation} \label{GV}
\frac{v_g}{c}\!= \!\left\{1- \frac{c^2}{2\omega_0^2}\left[  C_{\omega_0}(z)-\omega_0 C^{\prime}_{\omega_0}(z) \right] \right\}^{-1} ,
\end{equation}
can be readily found. The values of $v_p$ and $v_g$ at a given cross section $z$ of a pulsed beam can be then predicted from the transversal profile of the monochromatic light beam at $\omega_0$ and neighbouring frequencies. It is worthwhile to note that the above expressions of $v_p$ and $v_g$ can be formally obtained from the well-known formulas $v_p/c=n_{\omega_0}^{-1}$ and $v_g/c= (n_{\omega_0} + \omega_0 n^{\prime}_{\omega_0})^{-1}$ for a plane pulse in a material medium with refraction index $n_\omega(z) = 1+ (c^2/2\omega^2)C_\omega(z)$.

%%%%%%%%%%%%%%% superluminality %%%%%%%%%%%%%%%%%%%%%

In the most frequent and interesting case of $C_{\omega}(z)<0$ (maximum amplitude on-axis),  the condition of superluminality [$v_g(z)>c]$ is found to be, from Eq. (\ref{GV}),
$|C_{\omega_0}(z)|' > |C_{\omega_0}(z)|/\omega_0$. Negative $v_g(z)$ can even take place if
$|C_{\omega_0}(z)|'> |C_{\omega_0}(z)|/\omega_0 + 2\omega_0/c^2$.
In terms of the equivalent refraction index, superluminality corresponds to $n'_{\omega_0}<(1-n_{\omega_0})/\omega_0\gtrsim 0$, i.e., to weak normal dispersion or anomalous dispersion, and negative $v_g$ corresponds to $n'_{\omega_0}< -n_0/\omega_0<0$, i.e., to strong anomalous dispersion.
It appears, in conclusion, that the only requirement to produce a desired superluminality is to introduce a suitable variation with frequency of the concavity of the transversal profile of the monochromatic components. These variations can be easily generated in laboratory using simple optical elements as lenses, or suitably designed filters:

%%%%%%%%%%%%%%%%%% FOCUSING %%%%%%%%%%%%%%%%%%%%%

Consider an arbitrary pulsed beam $E(r,t)$ at the entrance plane of a lens whose focal length is $f$, and let write its spectrum in the convenient form $\hat E_\omega(r)=\hat P_\omega b_\omega (r)$, where  $\hat P_\omega$ can be identified with the pulse spectrum at a typical point on the lens (e.g., $r=0$). The dependence of $b_\omega(r)$ with $\omega$ accounts for possible changes in the transversal amplitude and phase profile with frequency. At the focal plane, the spectrum amplitude is, as is well-known, $a_\omega \propto |\tilde b_\omega(\omega r/2\pi c f)|$, where
\begin{equation}
\tilde b_\omega(\rho) =2\pi \int_0^\infty dr r b_\omega(r)J_0(2\pi \rho r)
\end{equation}
is the two-dimensional Fourier transform of $b_\omega(r)$, $J_0(\cdot)$ is the zero-order Bessel function of first class, $\rho=(\xi^2+\eta^2)^{1/2}$, $(\xi,\eta)$ being conjugate variables of $(x,y)$. The concavity at the focus can be shown to be $|C_\omega|=(\omega/2\pi c f)^2 |(\tilde \Delta_\perp |\tilde b_\omega|/|\tilde b_\omega|)_{\rho=0}|$, with $\tilde \Delta_\perp \equiv \partial_{\xi\xi}+\partial_{\eta\eta}$.  Elemental properties of the Fourier transform lead to the relationship $|(\tilde \Delta_\perp |\tilde b_\omega|/|\tilde b_\omega|)_{\rho=0}|=4\pi^2\langle r^2\rangle_\omega$, where $\langle r^2\rangle_\omega \equiv \left[\int_0^\infty dr r^3 |b_\omega(r)|\right]/\left[\int_0^\infty dr r |b_\omega(r)|\right]$ is the squared rms width of the transversal profile amplitude $|b_\omega(r)|$ on the lens. In conclusion, $|C_\omega|=\langle r^2\rangle_\omega (\omega^2/c^2f^2)$ at the focus.

In the case of an amplitude profile $|b(r)|$ independent of frequency, the proportionality $|C_\omega|\propto \omega^2$ due to the focusing process makes condition $|C_{\omega_0}|'>|C_{\omega_0}|/\omega_0$ of superluminality to be satisfied.  Indeed, from Eq. (\ref{GV}), the simple result $v_g/c=(1-\langle r^2\rangle/2f^2)^{-1}\gtrsim 1$ for the group velocity at the focus is obtained.
The Gaussian case $b(r)=\exp(-r^2/s^2)$ with plane pulse front and in the limit of infinite Fresnel number $N_{\omega_0}=s^2/\lambda_{\omega_0}f$ ($\lambda_{\omega_0}=2\pi c/\omega_0$) was considered in Ref. \onlinecite{ZBOR}. We see, however, that the superluminality at the focus is  independent of the transversal amplitude and phase profiles, and of $N_{\omega_0}$.

Strong superluminality ($v_g\gg c$, $v_g<0$) may be obtained if the transversal profile $b_\omega(r)$ on the lens depends on frequency.  In this case the concavity $|C_\omega|$ at the focus varies with $\omega$ due to focusing ($\propto\omega^2$) and also to the possilbe variation of the width $\langle r^2\rangle_\omega$ on the lens with $\omega$. From Eq. (\ref{GV}) the group velocity at the focus is  given by
\begin{equation}\label{GV3}
\frac{v_g}{c}=\left(1- \frac{1}{2}\frac{\langle r^2\rangle_{\omega_0}}{f^2}-\frac{\omega_0}{2}\frac{\langle r^2\rangle'_{\omega_0}}{f^2}\right)^{-1} ,
\end{equation}
which is (as in the previous case) independent of the exact shape of $b_\omega(r)$, and also of its possible distortion when widening with frequency. Superluminality is enhanced with respect to the case of $\omega$-independent profile if $\langle r^2\rangle_\omega$ at the lens grows with $\omega$ (i.e., $\langle r^2\rangle'_{\omega_0}>0$)

Growth of width with frequency must be obtained making use of the spatially dispersive properties of  certain optical systems. A fair possibility is the use of graded mirrors or transparencies \cite{BE02} made of a stack of dielectric layers, \cite{PI96} like the rather simple one shown in Fig. \ref{filter}a. Such a device is designed to produce a certain transversal profile $|b_{\omega_R}(r)|$ of a certain width $\langle r^2\rangle_{\omega_R}$ when illuminated with a specific frequency $\omega_R$. When illuminated with neighboring frequencies, however, we have found that the produced profile changes, and in particular its spot size grows significantly with frequency. Fig. \ref{filter}b shows the frequency-dependent width of the output amplitude profile when the multilayer is illuminated with Gaussian beams of different frequencies but constant width.

Given $\omega_0$, $\langle r^2\rangle_{\omega_0}\gg\lambda_{\omega_0}^2$, and $\langle r^2\rangle'_{\omega_0}$ produced by a spatially dispersive system [as those of Fig. \ref{filter}b in the case of the graded mirror], the focal length to be used to obtain $v_g=\infty$ at the focus is, from Eq. (\ref{GV3}),
\begin{equation}\label{FOCAL}
f^2=(\langle r^2\rangle_{\omega_0} +\omega_0\langle r^2\rangle'_{\omega_0})/2 \;\;\; \mbox{($v_g=\infty$)}
\end{equation}
(slightly smaller $f$ will yield a negative group velocity). This focal length must moreover verify, for paraxiality, $f\gg \sqrt{\langle r^2\rangle}_{\omega_0}$, or from (\ref{FOCAL}),
\begin{equation} \label{MAIN}
\langle r^2\rangle'_{\omega_0} \gg \langle r^2\rangle_{\omega_0}/\omega_0 ,
\end{equation}
which can be regarded as the condition for the device in order to obtain strong superluminality. In the case of the multilayer of Fig. \ref{filter}, for the selected $\omega_0=3.409$ fs$^{-1}$, we obtained $\langle r^2\rangle_{\omega_0}=0.17$ mm$^2$, and $\langle r^2\rangle'_{\omega_0}
=1.645$ mm$^2$ fs. Then condition (\ref{MAIN}) is loosely satisfied ($\langle r^2\rangle'_{\omega_0}\simeq 33 \langle r^2\rangle_{\omega_0}/\omega_0$). The focal length for $v_g=\infty$ is in fact $f=1.7$ mm $\gg$ $\sqrt{\langle r^2\rangle_{\omega_0}}=0.41$ mm.

Due to the appreciable variation of the width with frequency at the lens plane imposed by Eq. (\ref{MAIN}),  we must ensure that not only the monochromatic component at $\omega_0$, but at any $\omega$ within the frequency band $2\Delta\omega$, focuses paraxially, i.e.,  $\lambda_\omega\ll \sqrt{\langle r^2\rangle}_\omega\ll f$ for all $\omega$ in $2\Delta\omega$.
If we assume, for simplicity, the approximate linear variation of the width $\sqrt{\langle r^2\rangle}_\omega\simeq \sqrt{\langle r^2\rangle}_{\omega_0}+\sqrt{\langle r^2\rangle}^{\,\prime}_{\omega_0} (\omega-\omega_0)$ within $2\Delta\omega$, we must then require $\sqrt{\langle r^2\rangle}_{\omega_0-\Delta\omega}\gg\lambda_{\omega_0-\Delta\omega}\simeq \lambda_{\omega_0}$ for the smallest width in the spectrum, and $\sqrt{\langle r^2\rangle}_{\omega_0+\Delta\omega}\ll f$ for the largest one. These two conditions are seen to be satisfied if
\begin{equation}\label{BAND}
\Delta \omega \ll \sqrt{\langle r^2\rangle}_{\omega_0}/\sqrt{\langle r^2\rangle}^{\,\prime}_{\omega_0}=2\langle r^2\rangle_{\omega_0}/\langle r^2\rangle'_{\omega_0} .
\end{equation}
Eq. (\ref{BAND}) then determines the spectral bandwidth of the input pulse to achieve strong superluminality by paraxial focusing. For the proposed filter, Eq. (\ref{BAND}) yields $\Delta\omega\ll 0.206$ fs$^{-1}$. In practice, the entire spectrum should lie within the operation range $\omega>\omega_R$ of the filter; one can then take, e.g., $\Delta\omega\le 0.206/16$, which implies an input pulse duration $\Delta t\ge 155$ fs.

To investigate in more detail the effects associated to infinite or negative $v_g$, we have simulated the focusing of the pulsed beam $E(r,t)$ whose spectrum $\hat E_\omega(r)=\hat P_\omega \exp(-r^2/s_{L,\omega}^2)$
has a frequency-dependent width $s_{L,\omega}$. Using standard rules of Gaussian beam propagation \cite{SI}, the spectrum amplitude at any propagation distance $z$ beyond the lens is given by $a_\omega(r,z)=[s_{L,\omega}\hat P_\omega/s_\omega(z)] \exp[-r^2/s_\omega^2(z)]$, where $s_\omega(z)=s_{L,\omega}[(1-z/f)^2+(2cz/\omega s_{L,\omega}^2)^2]^{1/2}$ is the Gaussian width at distance $z$, and the spectrum phase is given by $\phi_\omega(r,z)=-\tan^{-1}[\pi N_\omega(z-f)/z]-\pi/2 +\omega r^2/2cR_{\omega_0}(z)$,  where $N_\omega=s_{L,\omega}^2/\lambda_\omega f$ is the Gaussian Fresnel number, and $1/R_{\omega_0}(z)=[s_{L,\omega}/s_\omega(z)][(1-z/f)/f + (2c/\omega s_{L,\omega})^2 z]$.  The group velocity along the $z$-axis turns out to be, from Eq. (\ref{GV}),
\begin{equation}
\frac{v_g(z)}{c}=\left\{1- \frac{s_{L,\omega_0}^2}{2f^2}\sigma \frac{(1-\xi)^2[1-(\xi\pi N_{\omega_0})^2]}{[1+(\xi\pi N_{\omega_0})^2]^2}\right\}^{-1}    \label{VGG}
\end{equation}
where $\sigma\equiv 1+2\omega_0 s'_{L,\omega_0}/s_{L,\omega_0}$ and
$\xi\equiv (z-f)/z$. At the focus ($\xi=0$), Eq. (\ref{VGG}) reduces to Eq. (\ref{GV3})
with the identification $s_{L,\omega_0}^2=\langle r^2\rangle_{\omega_0}$.
The curve $v_g(z)$ is shown in Fig. \ref{group} for typical sets of parameters. The form of the curve is solely determined by $N_{\omega_0}$.
A maximum is reached at an intermediate point between the focus $z=f$ and the waist position $z_0=f/[1+1/\pi^2N^2_{\omega_0}]$ of the Gaussian beam with frequency $\omega_0$. If  $N_{\omega_0}$ is large enough,  the asymmetry of the curve $v_g(z)$ with respect to $z=f$ becomes inappreciable.
The maximum enhances with increasing convergence angle $s_{L,\omega_0}/f$. Moreover, for given $s_{L,\omega_0}/f$ (as in the three curves in Figs. \ref{group}a and b), arbitrary enhancements are achieved by increasing $s'_{L,\omega_0}$.
For instance, in Fig.\ref{group}b, $\omega_0 s'_{L,\omega_0}/s_{L,\omega_0}=600$ (so that the lhs of (\ref{FOCAL})  is 2/3 the rhs). Then an extended region with negative $v_g$ appears around the focus. This region plays a similar role as the gas cell in the experiments of Ref. \onlinecite{WANG}.

The behavior of a pulse with such an unusual $v_g$ is more easily understood from Fig. \ref{peak}a. The solid curve shows the arrival time of the pulse peak (that travels at velocity $v_g$) at axial position $z$, as determined by the pulse front equation \cite{BO}  $t=[\omega z/c + \phi_\omega(0,z) ]'_{\omega_0}$. The slope of the curve yields the group velocity in units of $c$.
At $z_A$ and $z_B$, $v_g$ is infinite, and negative between them. As a consequence (as is apparent from the curve),
the pulse peak may arrive earlier at axial points behind the focus than at points before the focus, e.g., earlier at $z_B$ than at $z_A$.
 The infinite group velocity at $z_B$ can be associated to the fact that the pulse peak appears to arrive instantly at $z_B$ from $z_C$. Indeed, in the interval $(t_C,t_A)$, the pulse peak arrives at three positions at the same time.

Of course, the above interpretation for the behavior of the pulse peak is valid if the pulse temporal form does not experience significant deformation during propagation, in which case it is also valid for any temporal feature of the pulse.
In Fig. \ref{peak}b we have numerically verified that this is the case. The pulsed beam of spectrum on the lens $\hat E_\omega(r)=\hat P_\omega \exp(-r^2/s_{L,\omega}^2)$,
with $\hat P_\omega=\sqrt{\pi}(b/2)\exp[-b^2(\omega-\omega_0)^2/2]$ [i.e., $P(t)=\exp(-t^2/b^2)$]  was propagated behind the lens according to the Gaussian beam rules \cite{SI}, and the time-domain field was obtained from numerical integration of Eq. (\ref{UNO}), with $\omega_0=1.571$ fs$^{-1}$,  $s_{L,\omega_0}=0.096$ mm, $f=1.92$ mm, and $s'_{L,\omega_0}=36.67$ mm fs (to obtain, as in Figs. \ref{group}b and \ref{peak}a, the values $N_{\omega_0}=4$,  $s_{L,\omega_0}/f=0.05$ and $\omega_0 s^{\prime}_{L,\omega_0}/s_{L,\omega_0}=600$).  Condition (\ref{BAND}) is satisfied taking $\Delta\omega=1.309\times 10^{-3}$ fs$^{-1}$, or $b=2/\Delta\omega=1528$ fs.  Fig. \ref{peak}b shows the normalized pulse temporal form at  $z_A=1.861$ mm and $z_B=1.974$ mm (points A and B of Fig.\ref{peak}a).
There is no appreciable deformation in the pulse, which arrives at $z_B$ (after the focus) 109 fs earlier than at $z_A$ (before the focus).  At other distances $z$, we have also observed pulse form invariance. The arrival time of pulse peak at each selected $z$ is shown (small squares) in Fig. \ref{peak}a.

Fig. \ref{front} offers a more complete off-axis picture of the propagation. The pulse front surface $t=[\omega z/c +\phi_\omega(r,z)]'_{\omega_0}$, or surface formed by all points of equal peak time, is depicted at some instants of time.  When the convergent front advancing towards the focus intersects the axis at $z_C$, a new elliptical branch is born at position $z_B$ beyond the focus (curves 1), starting its motion with $v_g=\infty$.
At $t=6400$ (1 fs later than curve 3), the two branches join, transforming into a planar front that coincides with the focal plane and backpropagates at $v_g\simeq-2c$, and a nearly elliptical front around the focus. At longer times (curves 4-6), the situation is nearly reversed; in particular, a branch of the front implodes at $z_A$ with $v_g=\infty$.
In short, the pulse peak (or any other pulse feature) diverges from the axial point $z_B>f$ before it reaches the focal plane. Indeed, the converging front never reaches this plane, but dies out at $z_A<f$.

%%%%%%%%%%%%%%%%% CONCLUSION %%%%%%%%%%%%%%%%%%%%%

According to our analysis, strong superluminality in vacuum, including negative group velocities, could be experimentally demonstrated by focusing rather arbitrary pulses, under the only condition that the spot size varies with frequency appropriately. The required variation has been shown to be loosely attained with a multilayer graded mirror. Other devices are under study. Superluminality is here an effect of the dephasing between the different spectral components of the pulse due to the frequency dependence of Gouy's phase.

The authors thank A. Piegari for helpful discussions.

\begin{figure}
%\centerline{\psfig{figure=fig1.ps,width=7cm,clip=,angle=0}}
\caption{a) An all-dielectric multilayer high transmittance coating, obtained from a stack of quarter-wave ($\lambda_{\omega_R}=555$ nm, or $\omega_R=3.396$ fs$^{-1}$) layers of alternate high ($n_H=2.20$) and low index ($n_L=1.48$) on a transparent substrate ($n_S=1.48$), producing a radially variable transmittance.
b) The radius of the multilayer is $r=10$ mm. We used the simulation computer program of Ref. \onlinecite{PRO} to obtain the radial transmittance curve $t_\omega(r)$ for $\omega_R$ and higher frequencies. When the multilayer is illuminated with the frequency independent Gaussian profile $\exp(-r^2/s_0^2)$, $s_0=0.5$ mm, the output is $t_\omega(r)\exp(-r^2/s_0^2)$, whose width is depicted (open circles). For convenience we choose $\omega_0=3.409$ fs$^{-1}$, for which $\langle r^2\rangle_{\omega_0}= 0.17$ mm$^2$ and $\langle r^2\rangle'_{\omega_0}= 1.645$ mm$^2$ fs}
\label{filter}
\end{figure}

\begin{figure}
%\centerline{\psfig{figure=fig2.ps,width=7cm,clip=,angle=0}}
\caption{Group velocity of focused Gaussian pulsed beams as functions of propagation distance from the lens.  In all cases $N_{\omega_0}=4$, $s_{L,\omega_0}/f=0.05$. (a) $\omega_0 s'_{L,\omega_0}/s_{L,\omega_0}=0, 0.003$, for the solid and dashed curves respectively. (b) $\omega_0 s'_{L,\omega_0}/s_{L,\omega_0}=600$. }
\label{group}
\end{figure}

\begin{figure}
%\centerline{\psfig{figure=fig3.ps,width=7cm,clip=,angle=0}}
\caption{(a) Position of the pulse peak as it propagates, as predicted by $t=[\omega z/c+\phi_\omega(0,z)]'_{\omega_0}$ and numerically calculated. (b) Normalized pulse temporal form at propagation distances given. See the text for numerical values of the parameters.}
\label{peak}
\end{figure}

\begin{figure}
%\centerline{\psfig{figure=fig4.ps,width=7cm,clip=,angle=0}}
\caption{Pulse front at times a) 1: 6346 fs, 2: 6378 fs, 3: 6399 fs. b) 4: 6401 fs, 5: 6422 fs, 6: 6454 fs.}
\label{front}
\end{figure}

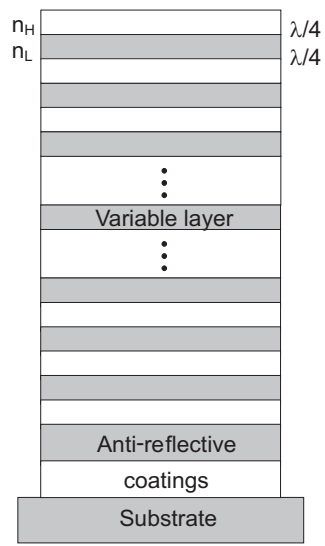
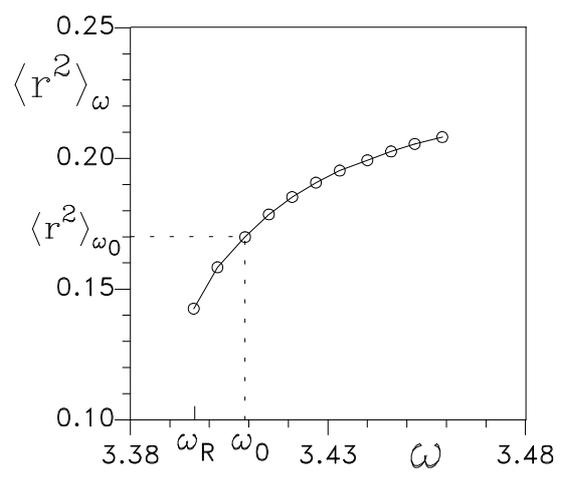
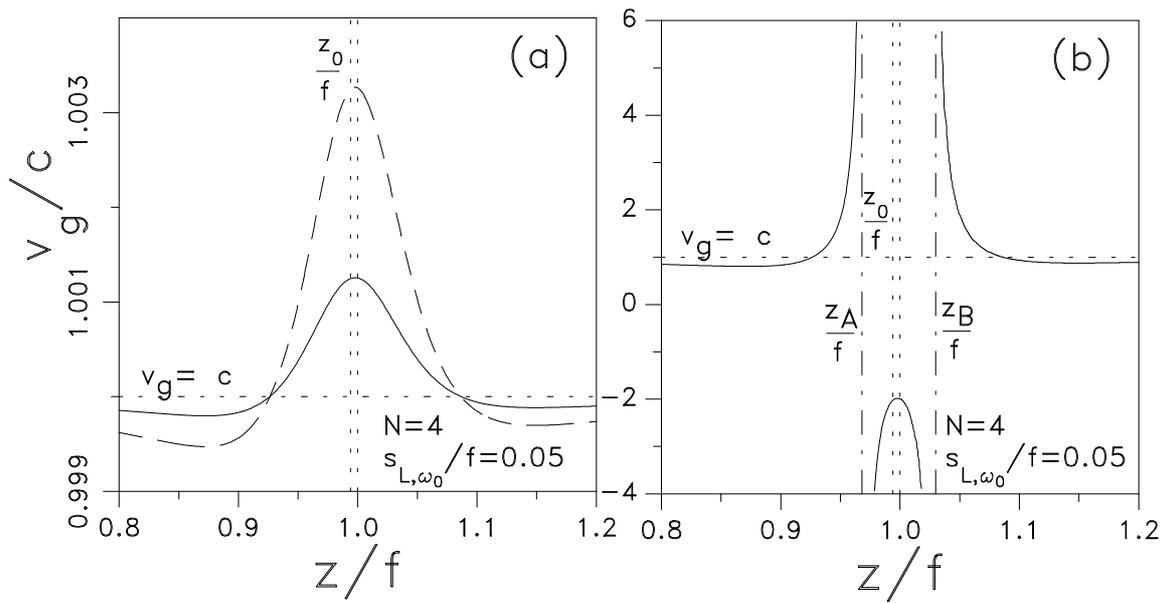
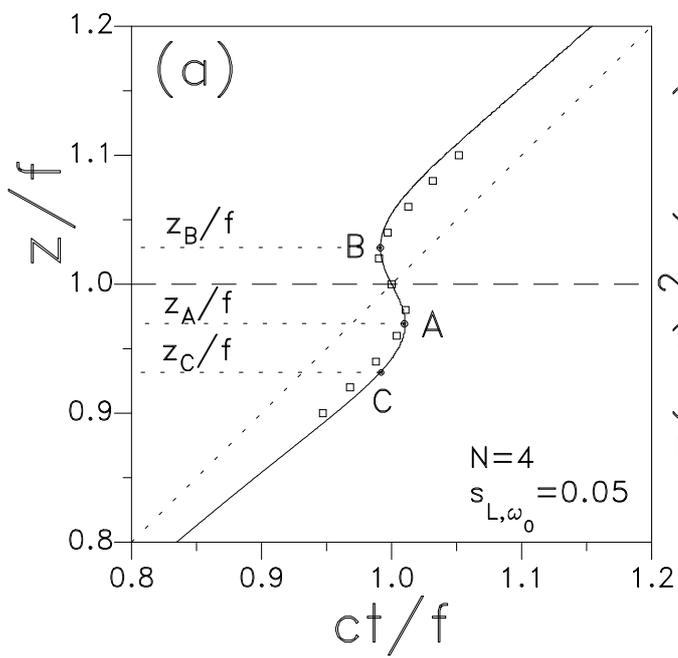
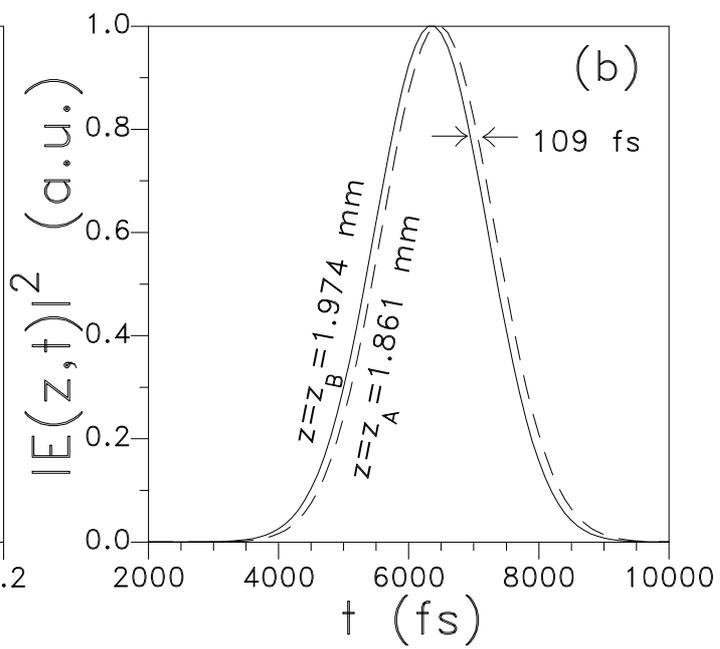
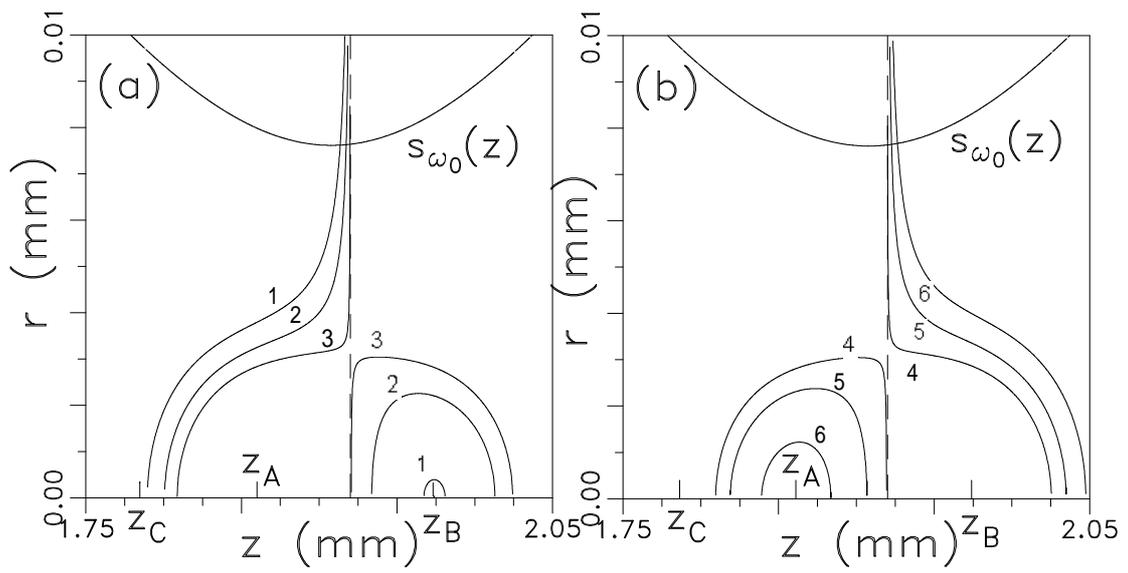
\end{document}